\documentstyle[preprint,pra,aps]{revtex}

\title{{\it Non Possum Comprimi Ergo Sum}: Skyrmions and Edge States in the
Quantum Hall Effect}
\author{A.H. MacDonald}
\address{Indiana University, Department of Physics, Bloomington, IN 47405, USA.}

\begin{document}
\draft
\preprint{IUCM96-002}

\date{\today}
\maketitle
\begin{abstract}
When the chemical potential of an electron system has a discontinuity at
a density $n^{*}$, the system is said to be incompressible and
a finite energy is required to create mobile charges
in the bulk of the system.  The quantum Hall effect
is associated with incompressibilities in a two-dimensional electron
system that occur at magnetic-field dependent densities, $n^{*}(B)$. In
these notes we discuss two aspects of the physics of quantum Hall systems
that follow directly from this association.
\end{abstract}

\section{Introduction}

The quantum Hall effect \cite{Klitzing82,qhebooks} is an anomaly that
occurs in the transport properties of two-dimensional electron system
(2DES) in the regime of strong perpendicular magnetic fields. At certain
magnetic fields it is found that the voltage drop in the system in the
direction of current flow, which is responsible for dissipation, vanishes
at low temperatures. Our understanding of this transport anomaly is not
absolutely complete, however there is fairly broad agreement that the
effect can occur only when the electronic system has, in the absence of
disorder, jumps in its chemical potential at certain densities ($n^{*}$),
that depend on magnetic field. In Section II of these notes we
specifically discuss the relationship between incompressibility and
transport properties in a 2DES. The remaining sections discuss two
aspects of the physics of quantum Hall systems that are direct
consequences of this relationship.

Quantum Hall systems can be ferromagnetic. Under appropriate
circumstances the spin-moments of the electrons align spontaneously, {\it
i.e.} in the absence of Zeeman coupling to a magnetic field. We will
refer to the 2DES in this case as a quantum Hall ferromagnet. Quantum
Hall ferromagnets provide a particularly simple example
of a two-dimensional itinerant
electron ferromagnet, and therefore represent an attractive target for
theories of the quantum statistical mechanics of such
systems \cite{kasner,read,ho}. They also have a number of
interesting unique properties,
the most striking of which is that their instantons carry an electronic charge.
This property is an immediate consequence of {\it incompressibility at a
magnetic-field dependent density}. In Section III of these notes we
discuss the electrically charged instantons of quantum Hall ferromagnets
and some of the observable consequences of their presence.

Another consequence of incompressibility at a magnetic field
dependent density is that quantum Hall systems {\it necessarily} have
gapless excitations localized at their edges. The low-energy physics of
quantum Hall systems is therefore like that of a one-dimensional electron
system which, because of time-reversal symmetry breaking by the magnetic
field, can carry a current even in equilibrium. In Section IV we will
discuss the physics of quantum Hall edges, including their description
in terms of the Luttinger liquid models appropriate to one-dimensional
fermion systems. Section V contains some concluding remarks.

These notes do not attempt a systematic introduction to the physics of the
quantum Hall effect. Interested readers can consult a previous effort of
mine \cite{leshouches} that is cited frequently below. If greater depth
is desired, readers can consult one of the excellent books covering
different aspects of the subject \cite{qhebooks}. I have written previously
at this level but at greater length on quantum Hall edge
states \cite{brazil}; parts of the present notes borrow from that
text.

\section{Incompressibility and the Quantum Hall Effect}

The thermodynamic compressibility of a system of interacting particles is
proportional to the derivative of the chemical potential with respect to
density. It can happen that at zero temperature the chemical potential has
a discontinuity at a density
$n^{*}$: the energy to add a particle to the system ($\mu^{+}$) differs,
at this density, from the energy to remove a particle from the system
($\mu^{-}$). The system is then said to be incompressible. In an
incompressible system a finite energy is required to create unbound
positive and negative charges that are capable of carrying current
through the bulk. The number of these free charges present in the system
will have an activated temperature dependence and will vanish for $T \to
0$. Incompressible systems are usually insulating. Paradoxically, as we
explain below, incompressibility is precisely the condition required for
the quantum Hall effect to occur. The twist is that in the case of the
quantum Hall effect, the density $n^{*}$ at which the incompressibility
occurs must depend on magnetic field. In my view, {\em incompressibility
at a magnetic-field-dependent density} is the {\it sine qua non} of the quantum
Hall effect.

For non-interacting electrons, the single-particle energy spectrum
of a 2DES in a magnetic field consists of Landau levels, separated by
$\hbar \omega_{c} \equiv \hbar e B / m^{*} c $ energy gaps and
with a macroscopic degeneracy $N_{\phi} = A B/ \Phi_{0} =
A / (2 \pi \ell^{2}) $. (Here $B$ is the magnetic field strength,
$ \Phi_{0} = hc/e $ is the magnetic flux quantum,
$A$ is the area of the system, and $m^{*}$ is the electron mass.)
Both the energy gaps between Landau levels and the degeneracy of the Landau
levels are proportional to $B$. Chemical potential discontinuities occur
whenever the density is an integral multiple of $n^{*} = B / \Phi_{0}$. We
show below that this property requires the existence of the gapless edge
excitations
discussed in Section IV. It is conventional in
discussing the quantum Hall effect to use a magnetic field dependent
density unit by defining the Landau level filling factor
$\nu \equiv n/n^{*} = 2 \pi \ell^{2} n$. For non-interacting
electrons, incompressibilities occur at integer filling factors.
When interactions are included, incompressibilities can also
occur at fractional filling factors and, as we will discuss,
physical properties near integer filling factors can be qualitatively
altered.

The relationship between incompressibility and the transport
anomalies that give the quantum Hall effect its name can
be understood by the following argument \cite{edgepicture}.
Consider a 2DES at zero temperature, as illustrated in
Fig.~[\ref{fig:incomp}].
We consider the case in which the chemical
potential lies in the `charge gap'; $ \mu \in (\mu^{-},\mu^{+})$. We want
to consider the change in the equilibrium local currents, present in the
system because of the breaking of time-reversal-invariance by the magnetic
field, when we make an infinitesimal change in the chemical potential,
$\delta \mu$. Because $\mu$ lies in the charge gap the change in the
local current density anywhere in the bulk of the system must be zero. The
current density can change, if it does anywhere, only at the edge of the
system. It follows from charge conservation that, if there is a change in
the current flowing along the edge of the system, it must be the same at
any point along the edge. We can relate this change in current to the
change in the orbital magnetization:
\begin{equation}
\delta I = {c \over A} \delta M.
\label{eq:orbmag}
\end{equation}
Eq.~(\ref{eq:orbmag}) is just the equation for the magnetic moment of a
current loop. However,
\begin{equation}
\delta M = {\partial M \over \partial \mu}\vert_{B} \, \delta \mu =
{\partial N \over \partial B}\vert_{\mu} \, \delta \mu.
\label{eq:maxwell}
\end{equation}
($N$ is the number of electrons.)
The second equality in Eq.~(\ref{eq:maxwell}) follows from a Maxwell
relation. Combining Eq.~(\ref{eq:orbmag}) and Eq.~(\ref{eq:maxwell}) we
obtain the following result for the rate at which the equilibrium edge
current changes with chemical potential when the chemical potential lies
in a charge gap:
\begin{equation}
{\delta I \over \delta \mu} = c {\partial n^{*} \over \partial B}|_{\mu}.
\label{eq:streda}
\end{equation}
The fact that $\delta I / \delta \mu \ne 0$ implies that whenever the
charge gap occurs at a density that depends on magnetic field, there {\em
must} be gapless excitations at the edge of the system. Properties
of these low-energy edge excitations are discussed in Section IV.

Eq.(~\ref{eq:streda}) is expected to apply to the edge states even when the
chemical potential lies only in a mobility gap and not in a true gap, as
illustrated schematically in Fig.~[\ref{fig:incomp}]. A net current can be
carried from source to drain across the system by changing the local
chemical potentials only at the edges and having different chemical
potentials along the two edges connecting source and drain. When bulk
states are localized, the two edges and the bulk are
effectively decoupled from each other. Eq.~(\ref{eq:streda}) then
also applies to transport currents, relating the current carried
from source to drain to the chemical potential
difference between the two edges, equal to
$e V_{H}$ where $V_{H}$ is the Hall voltage. There is no
voltage drop along an edge since each edge is in local equilibrium and
hence no dissipation inside the sample.
Eq.~(\ref{eq:streda}) is commonly known as
the St\v{r}eda-Widom formula \cite{streda}.
In using this picture to explain transport experiments in bulk systems it
is necessary to claim that the transport current will be carried entirely
at the edge of the system even when bulk states occur at the Fermi level,
as long as these states are localized. There are difficulties with this
argument as a complete explanation for all transport phenomena associated
with the quantum Hall effect, but that is another story and we will not
pursue it here.

\section{Quantum Hall Ferromagnets}

\subsection{Energy Scales}

The quantum Hall regime is usually understood as the regime in
which no qualitative change in physical properties results
from mixing of Landau levels by either interactions or disorder.
It is common in theoretical studies to truncate the Hilbert
space to a single orbital Landau level and include mixing
, if at all, only when making quantitative estimates for comparison with
experiment. The quantum Hall regime, then, assumes that
the Landau level separation, $\hbar \omega_{c}$ is larger than
other energy scales of interest. On the other hand
the ferromagnetic state is generally defined in terms of the properties of an
electronic system in the absence of a magnetic field. The term
{\it Quantum Hall Ferromagnet} appears to be an oxymoron.
To understand why it is not only sensible but also
of more than academic interest to add this category to our taxonomy
of electronic states,
it is necessary to consider the relevant energy scales for
the case of the semiconductor systems in which 2DES's are realized.
For a free-electron system in a magnetic field, the Zeeman
splitting of spin-levels $ g \mu_{B} B $ and the Landau
level separation $\hbar \omega_{c}$ are identical, apart from
small relativistic corrections. Electrons in states near the
conduction band minimum of a semiconductor behave like
free electrons \cite{bastard} except that band effects
renormalize the electron mass $m^{*}$ and the g-factor.
In the case of the GaAs systems, where the quantum Hall effect
is most often studied, band effects increase the Landau
level separation by a factor of $\sim 20$ and reduce the
Zeeman splitting by a factor of $\sim 4$. As a result for
typical experimental situations, the Landau level separation
(in temperature units) is $\approx 200 {\rm K}$, and the
characteristic scale for electron-electron interactions
is $\approx 100 {\rm K}$ while the Zeeman splitting is only
$\approx 2 {\rm K}$. We call a system a quantum Hall ferromagnet
if the electronic spins in the incompressible ground state
with density $n^{*}$ align in the absence of Zeeman coupling.
In many cases, the properties of a quantum Hall ferromagnet
with such a small Zeeman coupling do not differ noticeably
from the properties when the Zeeman coupling is set to zero.
In other cases, the small Zeeman coupling plays an important
role but it is still useful to treat the system as a ferromagnet
in the presence of a small symmetry breaking field.

\subsection{Ferromagnetic Ground States}

The Hartree-Fock approximation, in which
many-electron states are approximated by single Slater
determinants, provides a simple
explanation for itinerant electron ferromagnetism that
is often qualitatively correct. For non-interacting
electrons, energy is minimized by
occupying both spin-states of each single particle
energy level. (If the number of majority-spin electrons
exceeds the number of minority-spin electrons it is necessary
to occupy higher energy single-particle levels.) The ground state
thus has equal numbers of majority-spin and minority-spin electrons
if the total number of electrons is even and the difference is one
if the total number of electrons is odd. This statement
applies for an arbitrary spin-quantization
axis. As long as there is no spin-orbit coupling, the Hamiltonian
is invariant under global spin rotations and the total spin
of all electrons is a good quantum number.
The above statement is equivalent to the
observation that for non-interacting electrons the
ground state always has total spin $S =0$ if the number
of electrons is even and total spin $S =1/2$ if the
number of electrons is odd. However, interaction energies
are lower in single-Slater-determinant states with
higher values of the total spin and, generally, is
minimized in fully spin-polarized states with $S=N/2$.
As in the familiar Hund's rules from atomic physics,
higher spin states tend to have lower interaction energies because like-spin
electrons are prevented from being at the same position
by the Pauli exclusion principle and therefore have
more energetically favorable spatial correlations.
In the Hartree-Fock approximation,
or in closely related spin-density-functional
approximations, an itinerant electron system is expected to be
ferromagnetic if the reduction in interaction energy
due to creating a finite spin-polarization state exceeds the
cost in single-particle kinetic energy, or more generally
`band' energy. Because of Landau level degeneracy,
the cost in kinetic energy of creating a finite
spin polarization for electrons in a magnetic field is
precisely zero unless $\nu$ is an even integer.
Hartree-Fock or similar approximations would predict
a ferromagnetic ground state for electrons at nearly
any value of $\nu$. In fact, this conclusion is
incorrect. For example, at certain filling factors it is
known \cite{singlet} that the interaction energy is minimized in a $S=0$
state. We do believe, however, that there exist finite ranges of filling
factor over which the ground state has $S/N \ne 0$.
For a deeper understanding of this behavior, we need a more rigorous
argument.

The approach we now describe is in the same spirit as
the illuminating outlook on the spin-polarized fractional
quantum Hall effect that arises from appropriate hard-core model
Hamiltonians \cite{haldane,macdmur}. As discussed for the case of
interest below, these models have zero energy many-particle
eigenstates that are often known analytically, are separated from
other many-particle states by a finite gap, and have a
degeneracy that increases with decreasing $N$.
The incompressible state responsible \cite{leshouches}
for a quantum Hall effect transport anomaly in such a model
is the nondegenerate maximum $N$ zero energy eigenstate. The zero energy
eigenstates at lower densities constitute the portion of the
spectrum that involves only the degrees of freedom
of the, in general fractionally charged \cite{laughlin},
quasiholes of the incompressible state. It is assumed that
the difference between the model Hamiltonian and the true
Hamiltonian is a sufficiently weak perturbation
that the quasihole states are still
well separated from other states in the Hilbert space, although
accidental degeneracies will be lifted in the spectrum of the
true Hamiltonian. Here we apply this approach to argue
that the ground state at $\nu =1$ is a quantum Hall ferromagnet
with $S = N/2$.

For our analysis we use the symmetric gauge in which the
single-particle orbitals \cite{leshouches} in the lowest Landau level are
\begin{equation}
\phi_{m}(z) = \frac{z^{m}}{(2^{m+1} \pi m!)^{1/2}} \exp ( - |z|^{2}/4),
\label{eq:1}
\end{equation}
where \cite{remarknorb} $m = 0,1, \cdots, N_{\phi}-1$,
$z = x+ i y$, and $x$ and $y$ are the Cartesian components of the
two-dimensional coordinate. We study here a hard-core model for which the
interaction is:
\begin{equation}
V = 4 \pi V_{0} \sum_{i<j} \delta^{(2)}(\vec r_{i} - \vec r_{j})
\label{eq:ham}
\end{equation}
At strong magnetic fields the low-energy Hamiltonian is simply the
projection of this interaction onto the lowest Landau
level \cite{haldane}. Many-particle wavefunctions that are zero energy
eigenstates of this Hamiltonian must vanish when any two-particles are at
the same position and must therefore have the
difference coordinate for each pair of particles as a factor:
\begin{equation}
\Psi[z,\chi] = \big[ \prod_{i<j} (z_{i}-z_{j}) \big] \, \Psi_{B}[z,\chi].
\label{eq:map}
\end{equation}
We note that the each complex coordinate appears to the power
$N-1$ in the factor in square brackets in Eq.~(\ref{eq:map})
and that this factor is completely antisymmetric. It follows that
$\Psi_{B}[z]$ must be a wavefunction for $N$ {\it bosons} and that
these bosons can be in states with angular momenta from
$0$ to $N_{\phi}-N$. This simple observation leads to the
conclusions we reach below.

In these notes we discuss only the case where
$N = N_{\phi}$; the same approach can \cite{hcmskyrmion} be extended to
$N \ne N_{\phi}$ to elucidate the physics of charged excitations in
quantum Hall ferromagnets but here we will follow another
line for this part of the discourse. For $N = N_{\phi}$ all bosons must
be in orbitals with $m=0$. $\Psi_{B}[z,\chi]$ must then be
proportional to a symmetric many-particle spinor and therefore
have total spin quantum number $S=N/2$. The orbital part of
the fermion wavefunction can be recognized as the Slater
determinant with all orbitals from $m=0$ to $m=N_{\phi}-1$ occupied.
We are able to conclude
that the ground state is a strong ferromagnet with no
orbital degeneracy. The ease with which this conclusion can
be reached contrasts markedly with the case of the Hubbard model
where enormous effort has yielded relatively few firm
results \cite{hubbard}. When Zeeman coupling is included in the
Hamiltonian, the ground state will be the member of this $N+1$ fold
degenerate multiplet for which all spins are aligned with the magnetic
field, {\it i.e.} the state with $S_{z} =S =N/2$.

\subsection{Charged Instantons}

In this section we follow a line of argument that
emphasizes the role of
{\it incompressibility at a magnetic field dependent
density} in the unusual properties of quantum Hall
ferromagnets. For the most part we follow Sondhi {\it et al.},
in applying the non-linear $\sigma$ model ($NL\sigma$)
field-theoretical description of a ferromagnet
to the present case \cite{sondhi}. The $NL\sigma$ model is intended to
capture the long-wavelength low-energy physics of isotropic ferromagnets.
In the $NL\sigma$ model, the energy is expressed as a functional
of a unit vector $\hat m (\vec r)$, which specifies the direction
of the ordered spin-moment as a function of position:
\begin{equation}
E = E_{0} + \frac{\rho_{s}}{2} \int d^{2} \vec r \, |\nabla \hat m(\vec r)
|^{2} .
\label{eq:nlsigmod}
\end{equation}
We will refer to a particular configuration of the ferromagnet specified
by a function $\hat m(\vec r)$ as a spin-texture.
Here $E_{0}$ is the ground state energy, which is independent of the
direction of the ordered moment as long as it is constant in space,
and $\rho_{s}$, the spin-stiffness,
is a phenomenological constant that must be
determined from experiment or calculated from a microscopic
model. The low-energy long-wavelength physics of ferromagnets
is dominated by variations in the direction of the ordered
moment that are slow on microscopic length scales and
that can cost vanishingly small energies.

An important aspect of any field theory is the enumeration of its
instantons \cite{rajaraman}. Instantons are finite excitation energy
extrema of an energy (or action) functional, in which the variation
of the field is localized in space. Typically, the stability of
an instanton is associated with a topological classification of
field configurations.
For $NL\sigma$ models in two space dimensions and with three-dimensional
unit vector fields, all finite energy spin configuration can be
classified by an integer valued topological index,
sometimes called a topological charge,
which specifies the number of times the order parameter field is
wrapped around the unit sphere
when the position $\vec r$ is varied
over two-dimensional space. Field configurations with different
topological indices cannot be continuously deformed into one
another. The sign of the topological index
depends on the sense of the closed paths traced out
on the surface of the unit sphere when $\vec r$ traces a
closed path in space.  An explicit expression for
the topological index, $Q[\vec m]$, associated with a
spin-texture can be derived by first calculating the solid angle enclosed
on the sphere by $\hat m (\vec r)$ when $\vec r$ encloses an infinitesimal
area element in space, and then integrating over space:
\begin{equation}
Q[\vec m] = \frac{1}{4 \pi} \int d^{2} \, \vec r
\hat m (\vec r) \cdot [ \partial_{x} \hat m(\vec r)
 \times \partial_{y} \hat m(\vec r) ].
\label{eq:topcharge}
\end{equation}
Eq.[~\ref{eq:topcharge}] follows from the observation that
for a unit vector field,
$\partial_{i} \hat m(\vec r)$ is orthogonal to $\hat m(\vec r)$.

For the 2D $NL\sigma$ model it is possible to derive analytic expressions
for the lowest energy spin-textures of a given topological
charge \cite{rajaraman,originalrefs}. The lowest energy
textures with unit magnitude topological charge,
commonly called Skyrmions, have a spin-texture
of the following form:
\begin{eqnarray}
m_{x} & = & 2 x \lambda / (r^{2} + \lambda^{2}) \nonumber \\
m_{y} & = & \pm 2 y \lambda / (r^{2} + \lambda^{2}) \nonumber \\
m_{z} & = & (r^{2} - \lambda^{2}) / (r^{2} + \lambda^{2})
\label{eq:skyrmion}
\end{eqnarray}
Here $\lambda$ is an arbitrary length scale, $\hat m$
has an arbitrary global orientation fixed here by setting
$\hat m = \hat z$ for $r \to \infty$, and the skyrmion is
centered at an arbitrary point chosen as the origin of
the coordinate system. It is easy to verify that these
spin textures have topological charge $Q = \pm 1$ and
excitation energy $E - E_{0} = 4 \pi \rho_{s}$.
The form of the Skyrmion spin texture is illustrated in
Fig.[~\ref{fig:skyrmion}].

The $NL\sigma$ model considerations in the above paragraph are
appropriate for any 2D ferromagnet.
Skyrmions lead to contributions to the
physical properties of 2D ferromagnets that
have a non-analytic temperature dependence of the
form $ \exp ( - 4 \pi \rho_{s} / k_{B} T ) $ at low
temperatures. These non-analytic
terms are interesting, since they cannot be captured by
perturbative theories, but they typically produce
only subtle corrections to a temperature
dependence that is dominantly controlled by thermally excited
spin-wave excitations.
In the case of quantum Hall ferromagnets, however,
Skyrmions carry an electrical charge and as a consequence
play a more prominent role in determining the physical properties
of the system. To understand why Skyrmions carry a charge it is
useful to consider first the case of a non-interacting electron
in the presence of a magnetic field that couples to its orbital
degrees of freedom and
an independent strong Zeeman field that couples to its spin
and whose orientation,
specified by $\hat m (\vec r)$, varies
slowly in space. The direction of the electron spin will vary
in space to maintain alignment with the external magnetic field.
As is familiar from the calculation of spin Berry phases, the
changing spin direction changes the adiabatic
orbital Hamiltonian that now \cite{fradkin} takes the form:
\begin{equation}
H = \frac{1}{2 m} \big(\vec p + \frac{e}{c} (\vec A + \vec A_{B})\big)^{2}.
\label{eq:berryphase1}
\end{equation}
Here the effective magnetic field due to the varying spin
orientation,
\begin{equation}
\nabla \times \vec A_{B} = \frac{\Phi_{0}}{4 \pi}
\hat m \cdot \big[ \partial_{x} \hat m(\vec r) \times \partial_{y} \hat
m(\vec r) \big],
\label{eq:berryphase2}
\end{equation}
is proportional to the topological index density of the unit vector
field $\hat m(\vec r)$. In a Hartree-Fock approximation, the
exchange interaction will produce a strong effective magnetic
field that points in the same direction as the local order
parameter, so we can associate $\hat m(\vec r)$ above with the order
parameter field. It follows that the effective value of $N_{\phi}$
is changed by one when the order parameter has a
texture with unit topological index. Moreover, as we have emphasized
in these notes, the charge gap of the quantum Hall effect occurs
at an electron density that depends on magnetic field. For a
quantum Hall effect at Landau level filling factor $\nu$ we
conclude that the electron number at which the charge gap occurs
changes by $ \nu Q[\hat m]$ when the order parameter field has topological
index $Q[\hat m]$. The same conclusion can be
reached \cite{sondhi,kanelee} using the Chern-Simons Landau-Ginzburg theory
of the quantum Hall effect \cite{zhang,readgl} or by explicit
calculation \cite{moon}.

The Skyrmion excitations of quantum Hall ferromagnets
are responsible for striking physical effects because they
carry a physical charge and {\it are present in the ground
state} for filling factors near those at
which the incompressible state occurs. To be concrete,
we consider the quantum Hall ferromagnet that occurs
at $\nu =1$. For $N = N_{\phi}$ the ground state has $S=N/2$
as discussed above. For $N= N_{\phi} \pm 1$, the ground state
contains a single charged Skyrmion. The Skyrmion can be introduced
by changing the total electron number or, in what is the typical
experimental situation, by changing the magnetic field strength
and hence $N_{\phi}$. In the $NL\sigma$ model the energy of
a Skyrmion is independent of its size. To represent Skyrmions in
quantum Hall ferromagnets, however, it is necessary to add
additional terms to the model to account for Zeeman coupling
and for the Coulomb self-interaction energy of a Skyrmion.
Zeeman coupling favors small Skyrmions, since the spin
near the center of the Skyrmion is oriented in opposition to the
Zeeman field. On the other hand the repulsive Coulomb energy
favors large Skyrmions. For typical experimental situations
the optimal Skyrmion size is not so much larger than
microscopic lengths, invalidating the use of the field-theory
description for quantitative estimates. In a quantum description,
the number of reversed spins per skyrmion is quantized so
that, when Skyrmion-Skyrmion interactions can be
neglected, we expect that the component of the total spin
along the direction of the Zeeman field is
\begin{equation}
S_{z} = N/2 - (K + \theta) |N - N_{\phi}|
\label{eq:szqn}
\end{equation}
Here $\theta = 1$ for $N > N_{\phi}$ and $\theta =0$ for $N < N_{\phi}$.
$|N - N_{\phi}|$ is the number of Skyrmions or antiskyrmions present
in the system. The integer quantum number $K$ will depend in general
on the relative size of Zeeman and Coulomb interaction terms
and is the relevant quantum measure of the Skyrmion size. For
non-interacting electrons, or with interactions treated in the
Hartree-Fock approximation, $K=0$ so that $S_{z}$ always has the
maximum value allowed by the Pauli exclusion principle.
($K$ is guaranteed by particle-hole symmetry \cite{phsym} to have the same
value for $N > N_{\phi}$ and $ N < N_{\phi}$.)
The $NL\sigma$ model considerations of Sondhi et al. \cite{sondhi}
described above, and also earlier numerical exact
diagonalization calculations \cite{ednumerical}, suggest
that $K$ should be non-zero for quantum Hall ferromagnets
and quite large if the Zeeman energy is small.
These predictions were dramatically confirmed when
Barrett {\it et al.} unexpectedly succeeded \cite{barrett} in using optical
pumping techniques to perform NMR Knight shift measurements of the
spin-polarization of two-dimensional electron systems in the
quantum Hall regime. The results of this experiment are
illustrated in Fig.~\ref{fig:barrett}
and correspond to
$K=3$, in quantitative agreement with microscopic predictions
based on a generalized Hartree-Fock approximation for
single-Skyrmion states \cite{fertigskyrmion}.
There seems to be little doubt that the elementary charged
excitations of quantum Hall ferromagnets are Skyrmion-like
objects that carry large spin quantum numbers.
Recent transport \cite{andy} and optical \cite{goldberg}
experiments add additional support to this conclusion.

For large enough $| N - N_{\phi}| $ the Skyrmion-like
objects will eventually interact strongly. When
the density of Skyrmions is low and the temperature
is low, Skyrmions are expected to form a lattice
similar to the Wigner crystal state formed by
electrons in the limit of very strong magnetic fields.
In Fig.~\ref{fig:sklat}
we compare theoretical calculations of the spin-polarizations as
a function of filling factor for several candidate
Skyrmion lattice states with experimental data.
The theoretical results were obtained by Brey {\it et al.}
using a generalized Hartree-Fock approximation\cite{sklatpaper} and
illustrate
several important aspects of the physics of Skyrme crystals.
These authors find that the ground state of the Skyrme
crystal is a square lattice rather than a triangular lattice
as found for the electron Wigner crystal.
Furthermore, as illustrated in Fig.~\ref{fig:sklat},
the spin-polarization of the square lattice Skyrme
crystal is much smaller, for a given Zeeman coupling strength,
than for the lowest energy triangular lattice state.
The preference for a square lattice can be understood
qualitatively in terms of the $NL\sigma$ model description of
Skyrmion states. For that model Skyrmions are centered at
an arbitrary point, have an arbitrary size, and are invariant
under arbitrary global spin rotations. When Coulomb and Zeeman
energies are included the optimal Skyrmion size is fixed and the
spin moment must be aligned with the Zeeman field far from the
Skyrmion center. However, the energy of each Skyrmion is still
invariant under global rotations of the moment about
an axis aligned with the Zeeman field. For a Skyrme lattice the
relative values of these rotation angles must be adjusted to
minimize the total energy. It turns out that the interaction
energy between a pair of Skyrmions is reduced when
they have opposing orientations for the component of the
ordered moment perpendicular to the Zeeman field. This
arrangement allows the ordered moment orientation to
vary more smoothly along the line connecting Skyrmion centers.
The tendency toward opposing orientations is frustrated on
triangular lattice, hence the energetic preference for a square
lattice. The stronger short-range repulsive interaction
in the aligned orientation ferromagnetic lattice case,
results in smaller Skyrmions and therefore more spin-polarized
states. The spin-polarizations calculated for the opposing orientation,
square lattice case shown in Fig.~\ref{fig:sklat} appear
to be in excellent agreement with experiment over a wide range
of filling factors near $\nu =1$.

\section{Edge Excitations of an Incompressible Quantum Hall Fluid}
\subsection{Non-Interacting Electron Picture}

Throughout this section we will consider a disk geometry where electrons
are confined to a finite area centered on the origin by a circularly
symmetric confining potential,
$V_{\rm conf}(r)$. We have in mind the situation where $V_{\rm conf}(r)$
rises from zero to a large value near $r = R$, where $R$ is loosely
speaking the radius of the disk in which the electron system is confined.
We choose this geometry, for which the electron system has a single edge,
since we limit our attention here to the properties of an isolated quantum
Hall edge and will not discuss the physics of interaction or scattering
between edges \cite{twoedgecaveat}. In this geometry it is convenient to
choose the symmetric gauge for which angular momentum is a good quantum
number.  For $V_{\rm conf}(r)$ the
single-electron kinetic energy operator has the macroscopically
degenerate Landau levels separated by $\hbar \omega_{c}$ and in
each Landau level states with larger angular momentum are
localized further from the origin.
We recall from Eq.(~\ref{eq:1}) that wavefunctions with angular
momentum $m$ are localized \cite{leshouches} near a circle with radius
$R_{m} = \sqrt{2(m+1)} \ell$. (Note that for large $m$ the separation
between adjacent values of $R_{m}$ is $ \ell^{2} / R_{m} << \ell$.) In the
strong magnetic field limit
the confinement potential does not mix different Landau levels.
Since there is only one state with each angular momentum in each Landau
level the only effect of the confinement potential is to
increase the energy of the symmetric gauge eigenstates when $R_{m}$ becomes
larger than $\sim R$. The typical situation is illustrated schematically in
Fig.~[\ref{fig:nonintel}].
Here the $n=0$ and $n=1$ Landau levels are
occupied in the bulk and the chemical potential $\mu$ lies in the gap
$\Delta =
\hbar \omega_{c}$ between the highest energy occupied Landau level ($E = 3
\hbar \omega_{c} /2$) and the lowest energy unoccupied Landau level ($E = 5
\hbar \omega_{c}/2$).  In this section we are
interested only in the ground state and the low energy
excited states obtained by making one
or more particle-hole excitations at the edge.  We will discuss
only the simplest situation where a single Landau level crosses the
chemical potential at the edge of the system and the analogous {\em
single branch} situations in the case of the fractional quantum Hall
effect \cite{impatedge}. We will also neglect the spin degree of freedom
of the electrons, which figured so prominently in the previous section.

An important property of the ground state of the non-interacting electron
system in the case of interest, is that it remains an exact eigenstate of
the system (but not necessarily the ground state!) when interactions are
present. That is because the total angular momentum $K$ for this state
is
\begin{equation}
M_{0} = \sum_{m=0}^{N-1} m = N (N-1) /2
\label{nov1a}
\end{equation}
and all other states in the Hilbert space (truncated to the lowest Landau
level) have larger angular momentum \cite{ajp}. For large disks and total
angular momentum near $M_{0}$ the excitation energy of a non-interacting
electron state will be
\begin{equation}
\Delta E = \gamma M
\label{nov1aa}
\end{equation}
where $M \equiv K - M_{0}$ is the excess angular momentum and $\gamma$ is
the energy separation between single-particle states with adjacent angular
momenta and energies near the Fermi energy. $\gamma$ is related to the
electric field, $E_{\rm edge}$ from the confining potential at the edge of
the disk:
\begin{equation}
\gamma = e E_{\rm edge} \frac{d R_{m}}{d m}
 = e E_{\rm edge} \ell^{2} / R
\label{nov1b}
\end{equation}
This expression for $\gamma$ can be understood in a more appealing way. In
a strong magnetic field charged particles execute rapid cyclotron orbits
centered on a point that slowly drifts in the direction perpendicular to
both the magnetic field and the local electric field. For an electron at
the edge of the disk the velocity of this `E cross B' drift is $v_{\rm
edge} = c E_{\rm edge}/ B $. The energy level separation can therefore be
written in the form
\begin{equation}
\gamma = \hbar v_{\rm edge}/ R_{\rm edge} = h / T
\label{nov1c}
\end{equation}
where $T$ is the period of the slow drift motion of edge electrons around
the disk, in agreement with expectations based on semiclassical
quantization.

Since the excitation energy depends only on the angular momentum increase
compared to the ground state it is useful to classify states by $M$. It
is easy to count the number of distinct many-body states with a given
value of $M$ as illustrated in Fig.~[\ref{fig:fermioncounting}].
For
$M=1$ only one many-particle state is permitted by the Pauli exclusion
principle; it is obtained by promoting the ground state electron with
$m=N-1$ to $m=N$. For $M=2$, particle hole excitations are possible from
$m=N-1$ to
$m=N + 1$ and from $m=N-2$ to $m= N$. In general $M$ many-particle
states with excess angular momentum $M$ can be created by making a
single-particle hole excitation of the ground state. For $M \ge 4$
additional states can be created by making multiple particle-hole
excitations. The first of these is a state with two particle-hole
excitations that occurs at $M=4$ and is illustrated in
Fig.~[\ref{fig:fermioncounting}].

\subsection{Many-Body Wavefunction Picture}

We now discuss the edge excitation spectrum of interacting
electrons using a language of many-particle wavefunctions. For the case
of the integer quantum Hall effect we will essentially recover the
picture of the excitation spectrum obtained previously for non-interacting
electrons by counting occupation numbers. We could have used the
Hartree-Fock approximation and occupation number counting to generalize
these results to interacting electrons. However, the Hartree-Fock
approximation is completely at sea when it comes to the fractional case.
Discussions of the fractional edge using an independent electron language
can be comforting but can also be
misleading.  Nevertheless, we will see that there is a one-to-one
correspondence between the edge excitation spectrum for non-interacting
electrons at integer filling factors and the fractional edge excitation
spectrum.

Many-electron wavefunctions where all electrons are confined to the
lowest Landau level must be sums of products of one-particle wavefunctions
from the lowest Landau level. From Eq.~(\ref{eq:1}) it follows that
any $N$ electron wavefunction has the form
\begin{equation}
\Psi[z] = P(z_{1},\ldots ,z_{N})\; \prod_{\ell} \exp{(-|z_{\ell}|^{2}/4)},
\label{eq:nov10a}
\end{equation}
where we have adopted $\ell$ as the unit of length and
$P(z_{1},\ldots,z_{N})$ is a polynomial in the two-dimensional complex
coordinates. This property \cite{analytic} of the wavefunctions will be
exploited in this section. The first important observation is that since
$\Psi[z]$ is a wavefunction for many identical fermions it must change
sign when any two particles are interchanged, and therefore must vanish
as any two particles positions approach each other. Since
$P(z_{1},\ldots,z_{N})$ is a polynomial in each complex coordinate it
follows \cite{murray} that
\begin{equation}
P(z_{1},\ldots ,z_{N}) = \prod_{i<j} (z_{i} - z_{j})\; Q[z]
\label{eq:nov10b}
\end{equation}
where $Q[z]$ is any polynomial that is symmetric
under particle interchange. It is important to note that the total
angular momentum of all the particles ($K$) is just the degree of the
polynomial $P[z]$, {\em i.e.} the sum of the powers to which the
individual particle complex coordinates are raised. Since the total
angular momentum is a good quantum number the polynomial part of any
many-electron eigenstate will be a homogeneous degree polynomial, {\em
i.e.} all terms in the many-particle polynomial must have the same degree.
Additionally, the total angular momenta corresponding to a polynomial
that, as in Eq.~(\ref{eq:nov10b}), is the product of two polynomials is
the sum of the angular momenta associated with those polynomials.

\,From the discussion of Section III it is clear that for non-interacting
electrons in any monotonically increasing confinement potential, the
lowest energy state will be the state with the minimum total angular
momentum. In Eq.~(\ref{eq:nov10b}) that corresponds to choosing $Q[z]$ to
have degree zero, {\em i.e.} to $Q[z] \propto 1$. It is easy to verify
that the wavefunction when $Q[z]$ is a constant is in fact the Slater
determinant formed by occupying the single-particle states with $m=0,
\ldots, N-1$. For interacting electrons this state will remain the ground
state provided the confinement potential is strong enough to overcome the
repulsive interactions between electrons that favor states with larger
total angular momentum. When this is the ground state, low-energy
excited states with excess angular momentum
$M = K - N (N-1)/2$ are linear combinations of the states constructed by
choosing all possible symmetric polynomials \cite{stone} of degree $M$ for
$Q[z]$.

We'll discuss the enumeration of these polynomials in a moment but pause
now to explain how this analysis many be generalized to the case of the
fractional quantum Hall effect. We limit our attention here
\cite{macdedge} to the simplest fractional quantum Hall effects that
occur at Landau level filling factors $\nu =1/m$ for any odd integer $m$;
in some senses the $m=1$ case can be regarded as a special case of the
fractional quantum Hall effect. The physics of the chemical potential jump
that occurs at these filling factors was explained in the pioneering
paper of Laughlin \cite{laughlin,leshouches}. For $\nu < 1/3$, for example,
it is possible to find states in the Hilbert space in which pairs of
electrons are never found in a state with relative angular momentum equal
to one. This is the two-body state in which two electrons are closest
together. All many-particle states that avoid placing pairs in
this state will have low energy.
If this condition \cite{caveat} is satisfied,
\begin{equation}
P(z_{1},\ldots ,z_{N}) \equiv \prod_{i<j} (z_{i} - z_{j})^{3}\; Q[z]
\label{eq:nov10c}
\end{equation}
for any symmetric polynomial $Q[z]$. If the system has an abrupt edge the
ground state will have $Q[z] \equiv 1$ just as in the non-interacting
case.  (Note that this approach to identifying the ground state
and low-energy excitations is very much the same as that used
in Section III to identify the ground state for $\nu$ near one when the
electrons are not spin-polarized, except possibly spontaneously!)
The edge excitations correspond to the same set of symmetric
polynomials as in the $\nu =1$ case. In the case of a model system with a
short-range interaction and a parabolic confinement potential, it is easy
to place \cite{largem} the argument we have sketched above on firm ground.
It is known from numerical studies that the bulk chemical potential
discontinuity survives when the model interaction is replaced by the
realistic Coulomb interaction.  However, special care is required
in considering edge excitations in physically realistic systems with
long range interactions and the reader is warned that the simple
models discussed below may not always apply.  These issues are
discussed at greater length elsewhere \cite{brazil}.

The wavefunction
\begin{equation}
\Psi[z] \equiv Q[z] \prod_{l} \exp (- |z_{l}|^{2}/4)
\label{eq:nov10d}
\end{equation}
is a wavefunction for $N$ bosons in a strong magnetic field. Thus the
enumeration of the edge excitations in terms of symmetric polynomials
discussed above is equivalent to enumerating all many boson wavefunctions
with a given value of the total angular momentum. The boson angular
momentum
\begin{equation}
M = \sum_{m=0}^{\infty} m\; n_{m}
\end{equation}
where $n_{m}$ are the boson occupation numbers, is equivalent (for $\nu =
1/m$) to the excess angular momentum $M = K - m N (N-1)/2$ of the fermion
wavefunctions. In the state with $M=0$,
$n_{0}=N$, all other boson occupation numbers are zero, and
$Q[z]$ is a constant. In the boson language the ground state is a Bose
condensate. The lone state with $M=1$ has $n_{1}=1$,$n_{0}=N-1$; the
symmetric polynomial for this boson wavefunction is
\begin{equation}
Q[z] = z_{1} + z_{2} + \ldots z_{N}.
\label{eq:nov10e}
\end{equation}
For the integer $\nu =1$ case, it can be shown explicitly that the
corresponding many-fermion state is the $M=1$ state with a single
particle-hole excitation at the edge, discussed in Section III. The set
of excitations at general values of $M$ can be described equally well in
either fermion or boson languages. Some of the states that occur at
small values of $M$ are listed in Table~\ref{table1}.

In the parabolic confinement case the total energy depends only on the
excess total angular momentum: $\delta E = \gamma M$. The number of
many-boson states with total angular momentum
$M$, $g(M)$ can be calculated by considering a system of non-interacting
bosons with single-particle energy $\gamma m$ so that $E = \sum_{m}
(\gamma m) \cdot n_{m} = \gamma M$. The partition function is
\begin{equation}
Z = \sum_{M} g(M)\; e^{-\gamma M/k_{B}T} = \sum_{M} x^{M}\; g(M)
\label{eq:nov12a}
\end{equation}
where $x = e^{-\gamma /k_{B}T}$. For $N \to \infty$, the $m=0$ state acts
like a reservoir with chemical potential $\mu = 0$ so that the partition
function calculation can be done in the grand canonical ensemble. The
degeneracies $g(M)$ can be read off the power series expansion of the
partition function:
\begin{eqnarray}
\lefteqn{Z = \prod_{k=1}^{\infty} \frac{1}{1 - x^{k}} = (1 - x)^{-1} (1
- x^{2})^{-1} (1 - x^{3})^{-1} \ldots}\nonumber\\
 &=& (1 + x + x^{2} + x^{3} + \ldots ) (1 + x^{2} + x^{4} + \ldots ) (1 +
x^{3} + x^{6} + \ldots ) \ldots \nonumber\\
 &=& (1 + x + 2 x^{2} + 3 x^{3} + 5 x^{4} + 7 x^{5} + 11 x^{6} +
 15 x^{7} + 22 x^{8} + \ldots.
 \label{eq:nov12b}
\end{eqnarray}
For \cite{largem} large $M$ $g(M) \sim e^{\sqrt{\frac{2}{3}}\cdot\pi\cdot
M^{1/2}}$. The function $g(M)$ is well known to number theorists from the
theory of partitions \cite{partitions} in which it is known as the
partition function, not to be confused with the physics partition
function above! For
parabolic confinement potentials and short-ranged repulsive
interactions, the degeneracy of the edge excitations at a given excess
angular momentum is exact in both integer and fractional $\nu =1/m$
cases. For general confinement potentials and general electron-electron
interactions these degeneracies will be lifted. However, there is reason
to expect that in the thermodynamic limit excitations with $M \ll
N^{1/2}$ will be nearly degenerate. One way to see this is to use the
chiral Luttinger liquid picture of quantum Hall edges that we discuss in
the following section. This approach will allow us to do more than
enumerate excitations of the system and, in particular will enable us to
discuss the density-of-states for tunneling into the edge of a quantum
Hall system.

\subsection{Chiral Luttinger Liquid Picture}

The chiral Luttinger liquid picture \cite{wenreview} of quantum Hall
systems is an adaptation of the Luttinger liquid theory of
one-dimensional electron systems. We start this section with a brief
outline of the portion of that theory that we require. Readers in search
of greater depth should look elsewhere \cite{mahanll}. As in higher
dimensions, low excitation energies states in a one-dimensional fermion
system will involve only single-particle states near the Fermi
wavevector. Since the differences in wavevector among the relevant
states at a given Fermi edge are small, the excitations produced by
rearranging them occur on length scales that are long compared to
microscopic lengths. It is therefore reasonable to argue that the energy
density in the system at any point in space should depend only on the
local density of left-moving ($k < 0$) and right-moving ($k>0$)
electrons, $n_{L}(x)$ and $n_{R}(x)$:
\begin{equation}
E[n_{L},n_{R}] = E_{0} + \int dx\;
\left[\frac{\alpha_{LL}}{2}\; \delta n_{L}^{2}(x) +
\frac{\alpha_{RR}}{2}\; \delta n_{R}^{2}(x) + \alpha_{LR}\; \delta
n_{L}(x)\; \delta n_{R}(x)\right].
\label{eq:enlnr}
\end{equation}
It is, perhaps, not completely obvious that the density provides a
complete parameterization of the low-energy excitations, and
indeed in the fractional Hall case there are situations where the analog
of Eq.~(\ref{eq:enlnr}) is incorrect.\cite{modelnotcorrect}
Here $\alpha_{LL}$, $\alpha_{LR}$ and $\alpha_{RR}$ are determined by the
second derivatives of the energy per unit length with respect to $n_{L}$ and
$n_{R}$ for a uniform system and can be determined in principle by a
microscopic calculation.
$\delta n_{L}(x)$ and $\delta n_{R}(x)$ are differences of the density from
the ground state density. Note that we have as a convenience chosen the
chemical potential to be zero in dropping a term proportional to $ \int
dx (\delta n_{L}(x) + \delta n_{R}(x) ) $. We start by considering the case
where $\alpha_{LR} =0$ so that the left-moving electrons and right moving
electrons are decoupled. Focus for this case on the energy of the right
moving electrons. We Fourier expand the density and note that
\begin{equation}
\int dx\; \delta n_{R}^{2}(x) = \frac{1}{L} \sum_{q\neq 0} n_{-qR}.
n_{qR}
\label{eq:nov12c}
\end{equation}
so that the energy can be written in the form
\begin{equation}
E_{R} = E_{0} + \frac{\alpha_{LL}}{2L} \sum_{q\neq 0}
n_{-qR} n_{qR}.
\end{equation}

The energy above can be used as an effective Hamiltonian for low-energy
long-wavelength excitations. The simplification at the heart of
the Luttinger liquid theory is the observation that when the Hilbert
space is truncated to include only low-energy, long-wavelength excitations
(in particular when the number of left-moving and right-moving electrons
is fixed) Fourier components of the charge density do not commute. For
example consider the second quantization expression for $n_{qR}$ in terms
of creation and annihilation operators with $k > 0$:
\begin{equation}
n_{qR} = \sum_{k>0} c_{k+q}^{\dagger} c_{k}^{\phantom{\dagger}}.
\end{equation}

An example of the dependence of the effect of products of these operators
on the order in which they act is more instructive than the actual
algebraic calculation of the commutators. Note for example that
\begin{equation}
n_{-qR}\; |\Psi_{0}\rangle = 0
\label{eq:nov12d}
\end{equation}
where $q > 0$ and $|\Psi_{0}\rangle$ is the state with all right-going
electron states with $k < k_{F}$ occupied and all right-going states with
$k > k_{F}$ empty. (The alert reader will have noticed that this state of
`right-going' electrons corresponds precisely to the `maximum density
droplet' states that occur in the quantum Hall effect.) $n_{-qR}$
annihilates this state because there are no right-electron states with a
smaller total momentum than $|\Psi_{0}\rangle$. On the other hand for
$q = M 2 \pi / L$, $n_{qR}|\Psi_{0}\rangle$ yields a sum of $M$ terms in
which single-particle hole excitations have been formed in
$|\Psi_{0}\rangle$. For example, if we represent occupied states by solid
circles and unoccupied states by open circles, as in
Fig.~(\ref{fig:fermioncounting}), for $M=2$ we have
\begin{eqnarray}
n_{qR}\; |\Psi_{0}\rangle &=& |\ldots\bullet\;\bullet\;\circ\;\bullet |
\bullet\;\circ\;\circ\ldots\rangle\nonumber\\
 && + |\ldots\bullet\;\bullet\;\bullet\;\circ |
\circ\;\bullet\;\circ\ldots\rangle.
\label{eq:nov12e}
\end{eqnarray}
Each of the $M$ terms produced by $n_{qR}|\Psi_{0}\rangle$ is mapped back to
$| \Psi_{0}\rangle$ by $n_{-qR}$. Therefore
$n_{qR} n_{-qR} |\Psi_{0}\rangle = 0 $ whereas
$n_{-qR} n_{qR} |\Psi_{0}\rangle = M |\Psi_{0} \rangle$. The general form of
the commutation relation is readily established by a little careful
algebra \cite{mahanll}:
\begin{equation}
[n_{-q'R},n_{qR}] = \frac{qL}{2\pi}\; \delta_{q,q'}.
\label{eq:nov12f}
\end{equation}
This holds as long as we truncate the Hilbert space to states with a
fixed number of right-going electrons and assume that states far from the
Fermi edge are always occupied.

We can define creation and annihilation operators for density wave
excitations of right-going electrons. For $q > 0$
\begin{eqnarray}
a_{q} &=& \sqrt{\frac{2\pi}{qL}}\; n_{-qR}\\
a_{q}^{\dagger} &=& \sqrt{\frac{2\pi}{qL}}\; n_{qR}
\end{eqnarray}
With these definitions Eq.~(\ref{eq:nov12f}) yields
\begin{equation}
[a_{q'},a^{\dagger}_{q}] = \delta_{q,q'}
\label{eq:nov27a}
\end{equation}
so that the density waves satisfy bosonic commutation relations.
Also note that
\begin{eqnarray}
\mbox{}[ \hat M,a_{q}] &=& - \frac{qL}{2 \pi} a_{q} \\
\mbox{}[ \hat M,a^{\dagger}_{q} ] &=& \frac{qL}{2 \pi} a^{\dagger}_{q}
\label{eq:nov27b}
\end{eqnarray}
where $\hat M$ is the total angular momentum operator.
The contribution to the Hamiltonian from right-going electrons is
therefore
\begin{equation}
H_{R} = \sum_{q>0} \hbar\; vq\; a_{q}^{\dagger} a_{q}^{\phantom{\dagger}}
\label{eq:hamiltonian}
\end{equation}
where
\begin{equation}
v = \frac{\alpha_{RR}}{2\pi\hbar} = \frac{1}{2\pi L\hbar}\;
\frac{d^{2}E_{0}}{dn_{R}^{2}} = \frac{1}{2\pi\hbar}\;
\frac{d\mu_{R}}{dn_{R}}
\label{eq:velocity}
\end{equation}
At low-energies the system is equivalent to a system of one-dimensional
phonons traveling to the right with velocity $v$. In the limit of
non-interacting electrons
\begin{equation}
v = \frac{\hbar k_{F}}{m^{\ast}} \equiv v_{F}
\end{equation}
as expected.

Without interactions between left and right-moving electrons a Luttinger
liquid is quite trivial. In particular the ground state
($|\Psi_{0}\rangle$) is a single-Slater determinant with a sharp Fermi
edge. For one-dimensional electron gas systems the interesting
physics \cite{mahanll} occurs only when left and right-moving electrons
are allowed to interact. Most notably, arbitrarily weak interactions
destroy the sharp Fermi edge that is the hallmark of Fermi liquids and
that survives interactions in higher dimensions. In the case of quantum
Hall edges, however, the above restriction to electrons moving in
only one direction is not a temporary pedagogical device.
The model with only right
moving electrons discussed above can be taken over {\em mutatis mutandis}
as a model of the edge excitations for an electron system with $\nu =1$.
The role played by the one-dimensional electron density
is taken over by the integral of
the two-dimensional electron density along a line perpendicular to the
edge.   In this way we arrive at the same bosonized picture of
the ground state and low-lying excitations at the edge of a
quantum Hall system as we reached previously by arguing in terms
of many-particle wavefunctions.  The single boson states which appeared
there are replaced by
the states of the chiral phonon system which has modes with only one sign
of momentum and velocity.

For $\nu = 1$ the analysis applies whether or not the electrons interact.
We now turn our attention to a discussion of the fractional case. Do all
steps of the above discussion generalize? We can argue that if
we are interested only in low-energy long-wavelength excitations, the
energy can be expressed in the form
\begin{equation}
E = E_{0} + \frac{\alpha}{2L} \sum_{q\neq 0} n_{-q} n_{q}.
\end{equation}
As we comment later, this expression can fail at the edge of
fractional quantum Hall systems although it is appropriate for
$\nu = 1/m$. What about the commutator?
There is an important difference in the line
of argument in this case, since single-particle states far from the edge
of the system are not certain to be occupied. Instead the average
occupation number is $\nu =1/m$ and there are large quantum fluctuations
in the local configuration of the system even in the interior. However,
we know \cite{macdedge} from the discussion in terms of many-body
wavefunctions in the previous section that the low-energy excitations at
$\nu =1/m$ {\em can} be described as the excitations of a boson system,
exactly like those at $\nu =1$,
which suggests that something like Eq.~(\ref{eq:nov12f})
must still be satisfied when the Hilbert space is projected to low
energies. If we replace the commutator by its expectation value in the
ground state we obtain
\begin{equation}
[n_{-q'},n_{q}] = \nu \cdot \frac{q L}{2\pi}\; \delta_{q,q'}
\label{eq:commutator}
\end{equation}
which differs from Eq.~(\ref{eq:nov12f}) only through the
factor $\nu$. It seems clear for the case of $\nu = 1/m$ this replacement
can be justified on the grounds that the interior is essentially
frozen (but in this case {\em not} simply by the Pauli exclusion
principle) at excitation energies smaller than the gap for bulk
excitations. What we need to show is that Eq.~(\ref{eq:commutator})
applies as an operator identity in the entire low-energy portion of
the Hilbert space. Below, however, we follow a different line of argument.

Appealing to the microscopic analysis in terms of many-body wavefunctions
we know that the excitation spectrum for $\nu =1/m$ is equivalent to that
of a system of bosons. We conjecture that the commutator $[n_{-q'},n_{q}] =
\propto q \delta_{q,q'}$. To determine the constant of proportionality we
will require that the rate of change of the equilibrium edge current with
chemical potential be
$ e \nu / h$. From the edge state picture of the quantum Hall effect
discussed in Section II, it is clear that this is equivalent to requiring
the Hall conductivity to be quantized at
$\nu e^{2} /h$. Since our theory will yield a set of phonon modes that
travel with a common velocity $v$ it is clear that the change in
equilibrium edge current is related to the change in equilibrium density by
\begin{equation}
\delta I = ev\delta n.
\end{equation}
When the chemical potential for the single edge system is shifted
slightly from its reference value (which we chose to be zero) the grand
potential is given by
\begin{equation}
E[n] = E_{0} + \mu\delta n + \alpha\frac{(\delta n)^{2}}{2}
\label{eq:nov13a}
\end{equation}
Minimizing with respect to $\delta n$ we find that
\begin{equation}
\delta n = \frac{\delta \mu}{\alpha}
\label{eq:nov13b}
\end{equation}
so that
\begin{equation}
\frac{\delta I}{\delta\mu} = \frac{ e v }{\alpha}
\label{eq:nov13c}
\end{equation}
In order for this to be consistent with the quantum Hall effect ($ \delta
I = (e \nu /h) \delta \mu $) our theory must yield a edge phonon
velocity given by
\begin{equation}
 v = \frac{\alpha}{h} \cdot \nu.
\label{eq:nov13d}
\end{equation}
The extra factor of $\nu$ appearing in this equation compared to
Eq.~(\ref{eq:velocity}) requires the same factor of $\nu$ to appear in
Eq.~(\ref{eq:commutator}). We discuss below the qualitative changes in the
physics \cite{wen,wenreview} of fractional edge states
which are implied by this outwardly innocent numerical factor.

It is worth remarking that the line of argument leading to this specific
chiral Luttinger liquid theory of the fractional quantum Hall effect is
not completely rigorous. In fact we know that this simplest possible
theory with a single branch of chiral bosons does not apply for all
filling factors \cite{macdedge,wen,macdj}, even though (nearly) all steps
in the argument are superficially completely general. The reader is
encouraged to think seriously about what could go wrong with our
arguments. Certainly the possibility of adiabatically connecting all
low-energy states with corresponding states of the non-interacting
electron system, available for one-dimensional electron gases and for
quantum Hall systems at integer filling factors but not at fractional
filling factors, adds confidence when it is available. In our view, the
microscopic many-particle wavefunction approach that establishes a
one-to-one mapping between integer and fractional edge excitations (for
$\nu =1/m$!) is an important part of the theoretical underpinning of the
Luttinger liquid model of fractional Hall edges. Once
we know that the edge excitations map to those of a
chiral boson gas and that the
fractional quantum Hall effect occurs, it appears that no freedom is left
in the construction of a low-energy long-wavelength effective theory.

An important aspect of Luttinger liquid theory is the expression for
electron field operators in terms of bosons \cite{mahanll}. This
relationship is established by requiring the exact identity
\begin{equation}
[\rho (x),\hat{\psi}^{\dagger}(x')] = \delta (x-x')\;
\hat{\psi}^{\dagger}(x')\;
\label{eq:nov13a1}
\end{equation}
to be reproduced by the effective low-energy theory. This equation simply
requires the electron charge density to increase by the required amount
when an electron is added to the system. The electron creation operator
should also be consistent with Fermi statistics for the electrons:
\begin{equation}
\left\lbrace\psi^{\dagger}(x),\psi^{\dagger}(x')\right\rbrace =0 .
\label{eq:nov13a2}
\end{equation}
In order to satisfy Eq.~(\ref{eq:nov13a1}), the field operator must be
given by
\begin{equation}
\hat{\psi}^{\dagger}(x) = c e^{i\nu^{-1}\phi (x)}
\label{eq:nov13a3}
\end{equation}
where $d \phi(x) / dx = n(x)$ and $c$ is a constant that cannot be
determined by the theory. The factor of $\nu^{-1}$ in the argument of
the exponential of Eq.~(\ref{eq:nov13a3}) is required because of the
factor of $\nu$ in the commutator of density Fourier components that in
turn was required to make the theory consistent with the fractional quantum
Hall effect. When the exponential is expanded the $k-th$ order terms
generate states with total boson occupation number
$k$ and are multiplied in the fractional case by the
factor $\nu^{-k}$; multi-phonon terms are increased
in relative importance. It is
worth remarking \cite{wen} that the anticommutation relation between
fermion creation operators in the effective theory is satisfied only when
$\nu^{-1}$ is an odd integer. This provides an indication, independent of
microscopic considerations, that the simplest single-branch chiral boson
effective Hamiltonian can be correct only when
$\nu =1/m$ for odd $m$. Wen \cite{wenreview} has surveyed, using this
criterion, the multi-branch generalizations of the simplest effective
Hamiltonian theory which are possible at any given rational filling
factor. His conclusions are consistent with arguments \cite{macdedge}
based on the microscopic theory of the fractional quantum Hall effect.

Eq.~(\ref{eq:nov13a3}) has been carefully checked
numerically \cite{palacios} and appears to be correct. The $\nu^{-1}$
factor leads to predictions of qualitative changes in a number of
properties of fractional edges. The quantity that is most directly
altered is the tunneling density-of-states. Consider, for example, the
state created when an electron, localized on a magnetic length scale, is
added to the ground state at the edge of a $N-$ electron
system with $\nu =1/m$:
\begin{eqnarray}
\hat{\psi}^{\dagger}(0) |\Psi_{0}\rangle &\sim& \exp{\left(
-\sum_{n>0} \frac{a_{n}^{\dagger}}{\sqrt{n \nu}}\right)}
|\psi_{0}\rangle\nonumber\\
 &=& 1 + \frac{1\mbox{ phonon term}}{\nu^{1/2}} + \frac{2\mbox{ phonon
terms}}{\nu } + \ldots.
\end{eqnarray}
The tunneling density states is given by a sum over the ground and
excited states of the $N+1$ particle system:
\begin{equation}
A(\epsilon) = \sum_{n} \delta(E_{n} - E_{0} - \epsilon)
| \langle \Psi_{n} | \psi^{\dagger}(0) | \Psi_{0}  \rangle |^{2}
\label{eq:sfun}
\end{equation}
Because of the increased weighting of multiphonon states, which become
more numerous at energies farther from the chemical potential,
the spectral function is larger at larger $\epsilon -
\mu$ in the fractional case. An explicit calculation \cite{wen,wenreview}
yields a spectral function that grows like
$(\epsilon-\mu)^{{\nu}^{-1}-1}$. It is intuitively clear that the
spectral function should be small at low-energies in the fractional case
since the added electron will not share the very specific correlations
common to all the low-energy states. It is amazing that by simply
requiring the low-energy theory to be consistent with the fractional
quantum Hall effect we get a very specific prediction for the way in
which this qualitative notion is manifested in the tunneling density of
states.

\section{Acknowledgments}

The ideas discussed here have been shaped by
discussions with members of the condensed matter theory group at Indiana
University, especially M. Abolfath, S.M. Girvin, C. Hanna,
R. Haussmann, S. Mitra,
K. Moon, J.J. Palacios, D. Pfannkuche,
E. Sorensen, K. Tevosyan, K. Yang, and U. Z\"{u}licke.
Discussions with L. Brey, R. Cote, H. Fertig, M. Fisher, M. Johnson,
C. Kane, L. Martin, J. Oaknin,
C. Tejedor, S.R.-E. Yang and X.-G. Wen are also gratefully
acknowledged. The responsibility for surviving misapprehensions
rests with me. This work was supported by the National Science Foundation
under grant DMR-9416906.

\begin{figure}
\caption{A large but finite two-dimensional electron gas. In
panel (a) the chemical potential lies in a gap and the only low-energy
excitations are localized at the edge of the system. In panel (b) the
chemical potential lies in a mobility gap so that there are low-energy
excitations in the bulk but they are localized away from the edge. In
panel (c) a net current is carried from source to drain by having local
equilibria at different chemical potentials on upper and lower edges.}
\label{fig:incomp}
\end{figure}

\begin{figure}
\caption{Illustration of a Skyrmion spin texture.  At the center of the
Skyrmion $\hat m$ points in the down ($-\hat z$) direction.
Far from the center of a Skyrmion $\hat m$ points in the up ($\hat z$)
direction. Along a ray at angle $\theta$ in a
circular coordinate system defined with respect to the
Skyrmion center, $\hat m$ rotates about an axis in the
$(\sin (\theta)  - \cos (\theta)$ direction from $-\hat z$ to $\hat z$.
At fixed $r$ the $\hat x - \hat y$ projection of $\hat m$
has fixed magnitude and rotates by $\pm 2 \pi$ when the angular coordinate
winds by $\pm 2 \pi$. At $r = \lambda$, $\hat m$ lies entirely
in the $\hat x - \hat y$ plane.}
\label{fig:skyrmion}
\end{figure}

\begin{figure}
\caption{Knight shift measurements by Barrett {\it et al.}
of the spin polarization of a two-dimensional electron gas near filling
factor $\nu =1$. Here $S=A=K+1/2$ so that the experiment
is consistent with $K=3$ for this sample. The dashed line
in this figure shows the dependence of spin-polarization on
filling factor expected for non-interacting electrons and, in the
Hartree-Fock approximation, also for interacting electrons. The
spin-polarization is assumed to be proportional to the Knight shift of the
${}^{71}Ga$ nuclear resonance and to be complete at $\nu =1$. (After
Ref.~\protect\cite{barrett}}
\label{fig:barrett}
\end{figure}

\begin{figure}
\caption{Dependence of spin-polarization $P$ on filling factor
for Skyrme lattice states.   Here $g$ is the ratio of the
Zeeman energy to the characteristic interaction energy
$e^2/\ell$ and the values chosen are typical of experimental
systems.  The open and closed circles are experimental
results of Barrett {\it et al.}.  The legends indicate the
nature of the Skyrme lattice state: the SLA state is a
square lattice state with opposing Skyrmion orientations;
the TLF state is a triangular lattice state with aligned
Skyrmion orientations. (After Brey
{\it et al.} in Ref.~\protect\cite{sklatpaper})}
\label{fig:sklat}
\end{figure}

\begin{figure}
\caption{Schematic spectrum for non-interacting electrons
confined to a circular disk in a strong magnetic field. In the limit of
large disks the dependence of the energy on $m$ can usually be considered
to be continuous. The situation depicted has Landau level filling factor
$\nu =2$ in the bulk of the system.  The low-energy excitations are
particle-hole excitations at the edge of the system.}
\label{fig:nonintel}
\end{figure}

\begin{figure}
\caption{Non-interacting many electron eigenstates for small
excess angular momentum $M$ specified by occupation numbers for the
single-particle states with energies near the chemical potential $\mu$.
The vertical bars separate single-particle states with $\epsilon_{m} < \mu$
from those with $\epsilon_{m} > \mu$. A solid circle indicates that $n_{m}=1$
in both the ground state and in the particular excited state; a shaded
circle indicates that
$n_{m} = 1$ in the particular excited state but not in the ground state; an
empty circle indicates that $n_{m} =0$.}
\label{fig:fermioncounting}
\end{figure}

\begin{table}[htb]
\caption{Quantum occupation numbers in boson and fermion descriptions for
edge excitations with small excess angular momentum $M$. $g_{M}$ is the
number of states with excess angular momentum $M$. The fermion occupation
numbers are relative to the maximum density droplet state. Only
non-zero values are listed for both fermion and boson descriptions.
$L=N-1$ is the highest angular momentum that is occupied in the maximum
density droplet state.}
\label{table1}
\begin{tabular}{llll}
M & $g_{M}$ & Fermion Description & Boson Description\\
\tableline
1 & 1 & $ n_{L+1} = 1, n_{L} = -1$ & $n_{1} = 1$\\
2 & 2 & $ n_{L+2} = 1, n_{L} = -1; n_{L+1} = 1, n_{L-1} = -1$ &
$n_{2} = 1; n_{1} = 2$\\
3 & 3 & $ n_{L+3}=1,n_{L}=-1;n_{L+2}=1,n_{L-1}=-1$
& $n_{3} = 1; n_{2} = 1, n_{1} = 1;$\\
 & & $ n_{L+1}=1 n_{L-2} =-1$
& $n_{1} = 3$\\
4 & 5 & $ n_{L+4}=1, n_{L}=-1; n_{L+3}=1,n_{L-1}=-1
$& $n_{4}=1;n_{3}=1,n_{1}=1;n_{2}=2$ \\
 & & $ n_{L+2}=1,n_{L-2}=-1; n_{L+1}=1,n_{L-3}=-1$ &
$n_{2}=1,n_{1}=2; n_{1}=4$ \\
 & & $ n_{L+2}=1,n_{L+1}=1,n_{L}=-1,n_{L-1}=-1$& \\
\end{tabular}
\end{table}


\begin{references}
\bibitem{caveat} The antisymmetry requirement generally requires
$P[z]$ must be an odd function of $z_{i}-z_{j}$ for any $i$ and $j$ so that
$P[z]$ contains only odd terms in a Taylor series expansion of its
dependence on this difference coordinate. The coefficient of the
$(z_{i}-z_{j})^{1}$ term must vanish if the pair never has relative angular
momentum equal to one so that the leading term is $(z_{i}-z_{j})^{3}$. It
then follows from analyticity that each difference coordinate to the third
power is a factor of $P[z]$.
\bibitem{modelnotcorrect} The microscopic analysis of fractional
quantum Hall edges in Ref.~\cite{macdedge} shows that the
parameterization of the energy density at the edge in terms
of the linear charge density is successful only for the simple
case where $\nu =1/m$ that we have discussed in these notes.
\end{references}
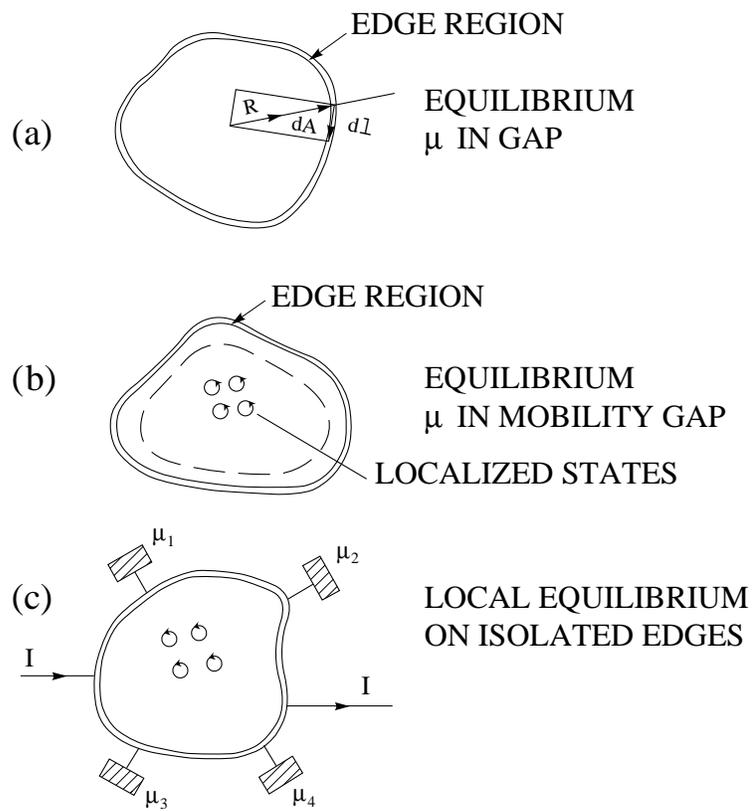
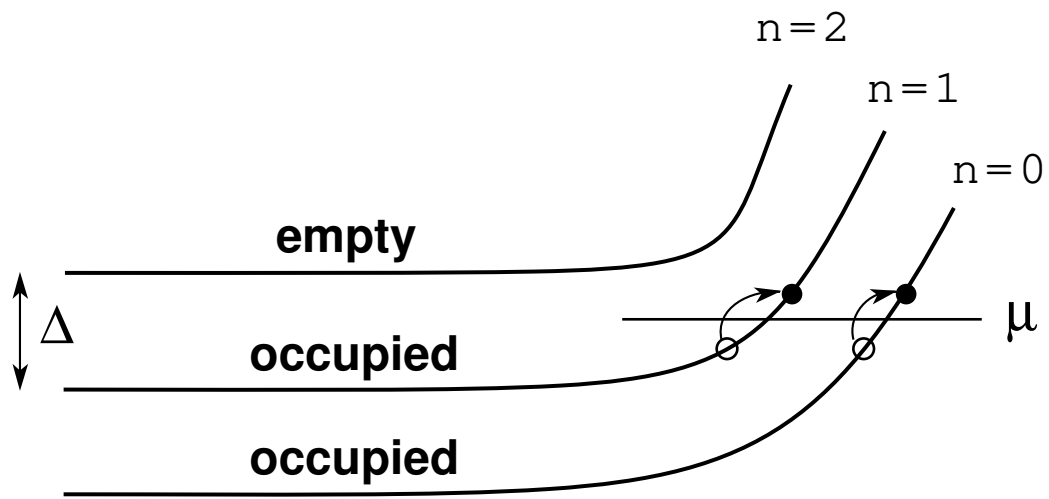
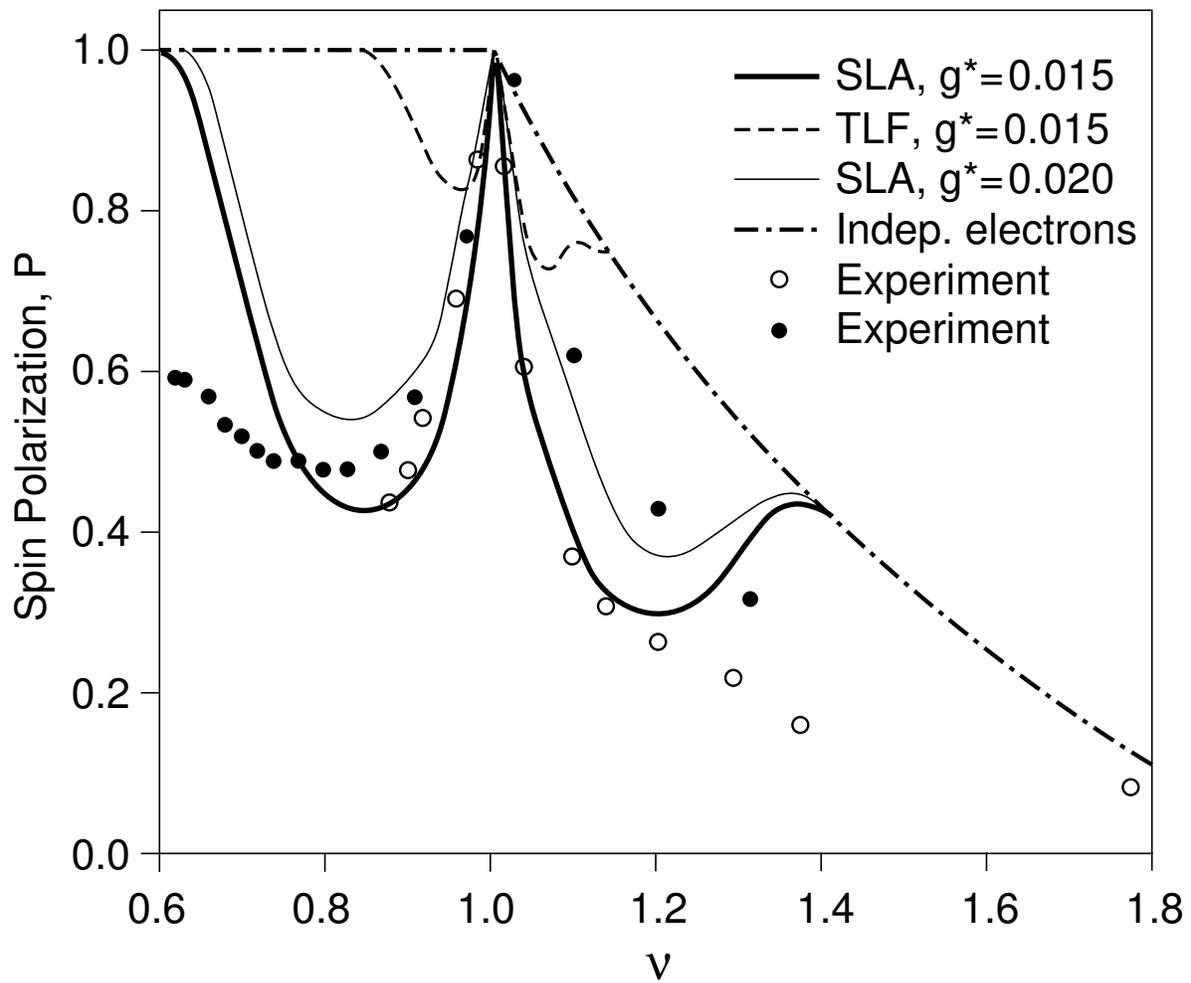
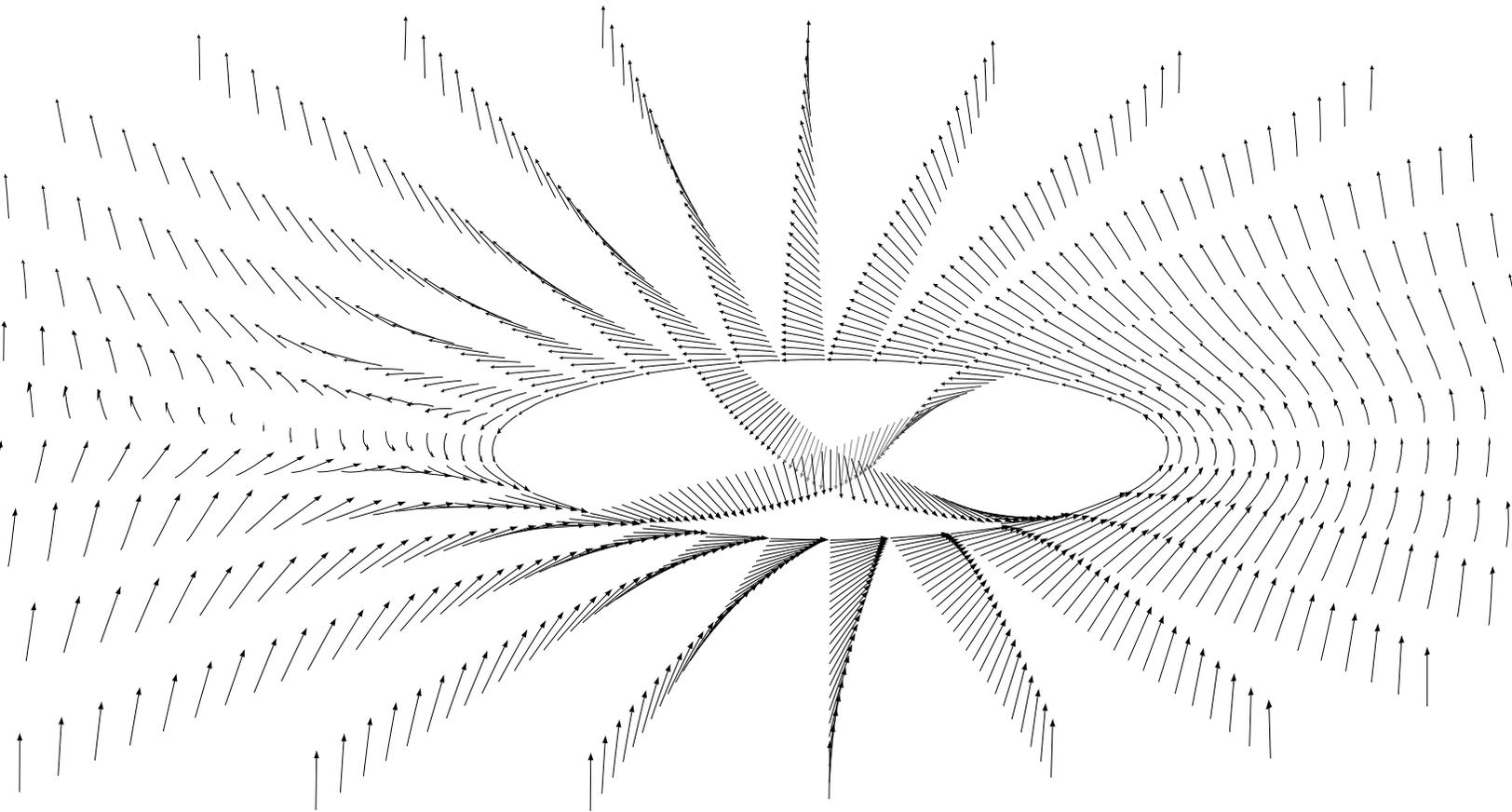
\end{document}